\documentclass[aps,prr,twocolumn,superscriptaddress,balancelastpage,longbibliography,10pt,reprint,nofootinbib]{revtex4-2}

\usepackage{graphicx}
\usepackage{amsmath,bbold}
\usepackage{times}
\usepackage{amssymb}
\usepackage{mathrsfs}
\usepackage{chemarr}
\usepackage{color}
\usepackage{url}
\usepackage{version}
\usepackage[linkcolor = blue, citecolor = blue, urlcolor = blue, colorlinks = true]{hyperref}
\usepackage{mwe,tikz}
\usepackage[percent]{overpic}
\usepackage[capitalise]{cleveref}
\usepackage{bm}
\usepackage[export]{adjustbox}
\usepackage{upgreek}
\definecolor{linkcolor}{rgb}{0,0,1}

\newcommand{\rr}{\bm{\mathrm{r}}}
\newcommand{\uu}{\bm{\mathrm{u}}}
\newcommand{\nn}{\bm{\mathrm{n}}}
\newcommand{\mm}{\bm{\mathrm{m}}}
\newcommand{\J}{\bm{\mathrm{J}}}
\newcommand{\QQ}{\bm{\mathrm{Q}}}

\usepackage{lipsum}

\usetikzlibrary{patterns}

\usepackage{ulem}
\begin{document}
  
%%%%%%%%%%%%%%%%%%%%%%%%%%

%\title{Noninteracting spherical active particles under confinement: generalized entropy potential}

\title{Active particles in a tube: a generalized entropy potential approach}

\author{Yongfeng Zhao}
\email{yfzhao2021@suda.edu.cn}
\affiliation{Center for Soft Condensed Matter Physics and Interdisciplinary Research \& School of Physical Science and Technology, Soochow University, 215006 Suzhou, China}
\affiliation{School of Physics and Astronomy and Institute of Natural Sciences, Shanghai Jiao Tong University, Shanghai 200240, China}

\date{\today}

\begin{abstract}
We study the transport of self-propelled noninteracting active
Brownian particles (ABPs) and run-and-tumble particles (RTPs) in long
tubes of varying widths. Using a moment expansion, we construct a
generalized Fick-Jacobs framework for the active particles when the
tube width is large and slowly varying. We show that the variation of
the particle density along the tube is well described by a
one-dimensional generalized entropy potential. This potential
resembles its passive counterpart, albeit with an effective
temperature and an effective tube width that are renormalized by the
activity. Our generalized entropy potential approach allows us to
predict the steady-state density distribution along the tube as well
as the mean escape time out of a spindle chamber. Finally, we show how
to account for the emergence of spontaneous ratchet flows in
asymmetric channel by including higher-order corrections neglected in
the effective entropy potential approach.
\end{abstract}

\maketitle

The transport of active particles through long tubes plays a critical
role in various biological and physical phenomena, particularly if the
tube's shape is irregular and its width changes~\cite{Wu_2015,
  Caprini_2019}. For example, channels with broken mirror symmetry
perpendicular to their axes can create spontaneous directed transport
of active particles, which is often referred to as ratchet transport
or ratchet flow~\cite{Leonardo_2010,angelani2011active, Ai_2014, Malgaretti_2017,reichhardt2017ratchet}. In terms of
biological implications, during the infection of a human body by
pathogens, microbes may migrate along the trachea, lymphangion,
urinary tract, or blood vessels. The question of transport along tubes
is also relevant to specific therapies, where engineered bacteria are
transported along blood vessels to tumors, stimulating the immune
system to kill cancer cells~\cite{Shi_2016}. Although real systems
often involve hydrodynamics and complex interactions, the study of
simple active-particle transport along tubes already constitutes a
rich and nontrivial starting point. Here we consider active Brownian
particles (ABPs) and run-and-tumble particles (RTPs)~\cite{Cates_2013,
  Solon_2015}, which are commonly used to describe the motion of
self-propelled colloids or microbes, such as bacteria~\cite{Berg_1972,
  Wilson_2011, Martinez_2012, Curatolo_2020}.

The transport of passive particles in channels has been extensively
studied~\cite{Jacobs_1967, Zwanzig_1992, Reguera_2001, Reguera_2006,
  Burada_2007, Yang_2017, Marbach_2018, Yang_2019, Zhu_2022}. In the
seminal work of Jacobs~\cite{Jacobs_1967}, it was shown that the time
evolution of the density of passive particles along a tube can be
reduced to the Fick-Jacobs (FJ) equation. The latter describes the
motion of particles in 2d or 3d tubes as an effective 1d motion along
the tube, subject to an effective potential proportional to the local
entropy of the particles, which is impacted by their motions in the
transverse directions. The reduction of dimensions simplifies the
analysis of phenomena such as first-passage problems and stochastic
resonance in tubes~\cite{Burada_2008, Ghosh_2010}.

On the contrary, the question as to whether active particles moving in
a long tube admit a similar framework has been little explored so
far. Unlike passive particles, active particles accumulate near rigid
walls~\cite{Elgeti_2013, Solon_2015_np, Duzgun_2018} due to their
persistent motion. As a result, the density of active particles along
a tube is not proportional to the area of its cross-section and the
implementation of a FJ approximation is thus harder than for passive
systems. Corrections accounting for the boundary accumulation---which
is proportional to the perimeter of the cross-section---indeed have to
be included. We expect these corrections to be particularly
significant when the particle persistence length is comparable to the
width of the tube.

So far, only the limit of narrow tubes---whose widths are comparable
to the particle size---has been discussed, in the presence of a strong
translational diffusion~\cite{Malgaretti_2017}. These results show
that a conservative potential cannot characterize the interplay
between the particle activity and the boundaries, and that asymmetric
channels can induce ratchet flows along the tube. What happens when
the tube width increases remains to be explored.

A relevant example where larger tubes have to be considered is that of
run-and-tumble bacteria \textit{Bacillus subtilis}. Indeed, the
persistence length of these bacteria is of the order of
$100\,{\rm\upmu m}$~\cite{Najafi_2019} and they can move in vessels
whose widths can be of the order of $40\sim 200\,{\rm \upmu
  m}$~\cite{Marbach_2018}. The bacterium size is typically a few ${\rm
  \upmu m}$ long~\cite{Ito_2005} and thus significantly smaller than
the vessel width. The translational diffusion of the bacterium is 3
orders of magnitude smaller than its effective large-scale
diffusivity, showing thermal noise to be largely irrelevant to
describe its large-scale transport~\cite{Ito_2005, Najafi_2019}. Note
that blood vessels can have lengths that are vastly larger than the
bacterium persistence length, which makes large-width long tubes
particularly relevant for the transport of the bacterium. The
transport of active particles in this regime has not been explored so
far.

In this article, we thus focus on the case of channels with
intermediate-to-large widths. We consider non-interacting ABPs and
RTPs moving in two or three dimensions in long channels of length $L$,
with rigid boundaries and varying width $w(x)$, where $\hat x$ is
along the tube axis. By `long channels', we refer to the case $L\gg
\ell_p$, where $\ell_p$ is the persistence length of the active
particles. We assume $w(x)$ is a slowly varying function such that
$|w'(x)|\ll 1$, which implies that the variations in widths over a
macroscopic length $L$ are negligible compared to $L$. In such a
regime, large-scale currents of particles provoked by the
inhomogeneity of confining walls are suppressed~\cite{Nikola_2016,
  Zakine_2020, BenDor_2022}. This allows us to treat the changes in
width perturbatively and to calculate analytically the time evolution
of the particle density and orientation fields along the tube. We show
that, in the limit of slow-varying width, the particle density along
the tube evolves in an effective entropy potential that we
characterize, while the orientation along the tube axis has the same
evolution as in a one-dimensional system in the absence of
confinement. The particle activity controls the effective temperature
of the system as well as the effective width entering the entropy
potential. Our formalism first allows us to predict the steady-state
density of particles along the tube as well as predict the
mean-first-passage time of a particle escaping from a spindle
channel. Finally, we show how a finite $|w'(x)|$ leads to a modified
evolution for the particle orientation field along the tube, which
allows us to quantitatively predict the ratchet flow in an asymmetric
channel. Throughout the article, our analytic calculations are
successfully compared with numerical simulations.

The article is organized as follows. In
Sec.~\ref{sec_FJ_prerequisites}, we review the FJ equation of passive
particles and outline the procedures and prerequisites which are
necessary for obtaining a FJ-like framework for active particles. In
Sec.~\ref{sec_FJ_active_0th}, we derive the joint dynamics of the
density and orientation field of the active particles along channels
of varying widths, which leads to a generalized entropy potential in
the limit of slow-varying width. We then show how the generalized
entropy potential can be used to predict the steady-state density
distribution of particles along the tube, how to generalize our
algebra to 3d systems, and how to calculate the mean-first-passage
time of a particle escaping from a spindle channel. In
Sec.~\ref{sec_FJ_active_1st}, we discuss how higher-order corrections
due to the variation of the channel width alter the dynamics of the
orientation field. This then allows us to predict the ratchet flow of
active particles in asymmetric tubes. Finally we close the article
with a brief discussion on elongated and interacting particles in
Sec.~\ref{sec_discussion}.

\section{The Fick-Jacobs equation}\label{sec_FJ_prerequisites}

In this section, we first review the Fick-Jacobs equation and the
concept of entropy potential for passive particles. Then we outline
the assumptions and prerequisites which lead to a Fick-Jacobs-like
framework for active particles.

\subsection{Passive particles and the Fick-Jacobs equation}\label{sec_FJ_passive}

Consider passive Brownian particles with diffusion constant $D$ moving
in a long channel oriented along the $x$-axis, with no-flux boundary
conditions along the $d-1$ other directions. We use arbitrary time and
length units. For tubes with varying width, the sectional area $A(x)$
of the channel is a function of the position $x$ along the tube. The
time evolution of the probability density function (PDF) $P(\rr,t)$ of
finding a particle at position $\rr$ at time $t$ satisfies the
diffusion equation
\begin{equation}
    \partial_t P=D\Delta P\;.
\end{equation}
On the tube surface, the system is subject to a no-flux boundary
conditions $\nn\cdot\nabla P=0$, where $\nn$ is the unit vector normal
to the surface.

The density of particles \textit{along the tube} can be found by
  considering the marginal distribution
  \begin{equation*}
  \rho(x,t)=\int_{\mathcal{S}(x)}
  P(\rr,t)d^{d-1}\rr\;,
  \end{equation*}
  where $\mathcal{S}(x)$ refers to the tube section of area $A(x)$.  If $A(x)$
  is a slowly varying function of $x$, the time evolution of
  $\rho(x,t)$ can be closed by assuming that the transverse dynamics
  along $\mathcal{S}(x)$ relaxes on time scales that are much shorter than the
  relaxation time of $\rho(x,t)$. The dynamics of $\rho(x,t)$ is then
  described by the Fick-Jacobs
  equation~\cite{Jacobs_1967,Zwanzig_1992}, which can be obtained by
  integrating the diffusion equation along all the spatial dimensions
  other than $x$, leading to:
\begin{equation}
\partial_t\rho=D\partial_{xx}\rho+\partial_x[\rho(x)V_{\rm eqm}'(x)]\; , \label{eqn_FJ_passive}
\end{equation}
where $V_{\rm eqm}(x)$ is a 1d effective potential defined as 
\begin{equation}
    V_{\rm eqm}(x)=-D\log A(x) \label{eqn_passive_potential}
\end{equation}
and referred to as an `entropy potential'. The physical intuition
behind Eq.~\eqref{eqn_passive_potential} is that particles spend more
time in wider regions, where the entropy potential is smaller.

\subsection{Generalized Fick-Jacobs framework for active particles}\label{sec_FJ_active_framework}
We now consider noninteracting ABPs in a long channel of length $L$ in
2d with a slowly varying width. The top and bottom boundaries of the
channel are described by $y=w_1(x)$ and $y=w_2(x)$, respectively. The width of the channel is then
$w(x)\equiv w_2(x)-w_1(x)$. Particles have constant speed $v$ and
persistence time $\tau$, and experience a translational diffusion with
diffusivity $D$. We denote by
$\uu(\theta)\equiv(\cos\theta,\sin\theta)$ the orientation of the
particle. The trajectory $\rr(t),\theta(t)$ of a single particle is
then given by the It\^{o}-Langevin equation
\begin{align}
&\dot{\rr}=v\uu(\theta)+\sqrt{2D}\bm{\xi}(t)\;, &\dot{\theta}=\sqrt{2\tau^{-1}}\eta(t)\;, \label{eqn_ABP_ito}
\end{align}
where $\bm{\xi}(t)$ and $\eta(t)$ are zero-mean Gaussian white noises
satisfying $\langle\xi_i(t)\xi_j(t')\rangle=\delta_{ij}\delta(t-t')$,
$\langle\eta(t)\eta(t')\rangle=\delta(t-t')$.  The PDF
$\Phi(\rr,\uu;t)$ of finding an ABP at position $\rr$ with orientation
$\uu$ then solves the Fokker-Planck (FP) equation:
\begin{equation}
\partial_t \Phi=-\nabla\cdot(v\uu\Phi)+D\Delta\Phi+\tau^{-1}\partial_{\theta\theta}\Phi\;. \label{eqn_ABP_FP}
\end{equation}
This is a continuity equation
$\partial_t\Phi=-\nabla\cdot\bm{\Gamma}-\partial_\theta\Gamma^\theta$,
from which we define the spatial flow of particles with orientation
$\uu$ as
\begin{equation}
    \bm{\Gamma}\equiv v\uu(\theta)\Phi(\rr,\uu;t)-D\nabla\Phi
\end{equation} 
and the angular flow as
$\Gamma^\theta=-(1/\tau)\partial_\theta\Phi$. The no-flux boundary
condition at the walls then simply reads $\nn\cdot\bm{\Gamma}=0$,
where $\nn$ is the normal vector to the boundary at $(x,w_{1,2}(x))$.

%We assume that the width changes slowly, over a macroscopic length scale, so that $|w'(x)|\sim \ll 1$, leading to width variations over the length $L$ that are much smaller that $L$. Technically, we will truncate Taylor expansions of $w(x)$ and build successive levels of approximations, depending on the order of the first derivatives of $w(x)$ that we neglect.

Let us consider the marginal distribution
\begin{equation}
\tilde{\Phi}(x,\uu)\equiv\int_{w_1(x)}^{w_2(x)}\Phi(x,y,\uu)dy\;.
\end{equation}
From now on, we use the tilde notation to denote the 1d marginal of
fields that describe their variation along the tube. Integrating
Eq.~\eqref{eqn_ABP_FP} over $y$ and using the no-flux boundary
condition leads to the exact time evolution:
\begin{align}
\partial_t \tilde{\Phi}=&-v\cos\theta\partial_x \tilde{\Phi}+(1/\tau)\partial_{\theta\theta}\tilde{\Phi}+D\partial_{xx} \tilde{\Phi} \nonumber \\
&-D\partial_x [\Phi(x,w_{2}(x),\uu) w_2'(x)-\Phi(x,w_{1}(x),\uu) w_1'(x)]\;. \label{eqn_FJ_Phi_noapprox}
\end{align}
As for passive systems~\cite{Jacobs_1967,Zwanzig_1992}, the key
assumption toward an FJ equation is $L\gg w(x)$, which implies that a
local steady state is reached in the $y$-direction at any $x$. To
close the equation for $\tilde{\Phi}(x,\uu)$ and obtain a generalized
entropy potential, we need to establish a connection between the
boundary values of the 2d PDF $\Phi(x,w_{1,2}(x),\uu)$ and the
marginal $\tilde{\Phi}(x,\uu)$.
  
Specifically, if we can solve the boundary values
$\Phi(x,w_{1,2}(x),\uu)$  as
\begin{equation}\label{eq:intermediate}
  \Phi(x,w_{1,2}(x),\uu)\simeq g(x,[w])\tilde{\Phi}(x,\uu)\;,
\end{equation}
then we can define a potential through
\begin{equation*}
  V'(x)=-D g(x,[w])w'(x)\;,
\end{equation*}
and rewrite the second line of Eq.~\eqref{eqn_FJ_Phi_noapprox} as
$\partial_x[\tilde{\Phi}(x,\uu)V'(x)]$. This would lead to an
effective entropy potential $V(x)$.

However, an explicit analytic solution of Eq.~\eqref{eq:intermediate}
is out of reach and we thus follow an alternative route, based on the
moment expansion of $\tilde{\Phi}$, that we detail in
Sec.~\ref{sec_FJ_active_0th}.

\subsection{Active particles with homogeneous confinement}\label{sec_ss_homo}

\begin{figure}
\begin{center}
\begin{tikzpicture}[line width=0.3mm, >=latex, scale=0.5]
\node at (-9,2.3) {\textbf{a)}};
\draw (-9,1.1) -- (-4,1.1);
\draw (-9,-3.1) -- (-4,-3.1);
\filldraw[fill=black] (3-11,-0.5) circle (0.03);
\draw[->] (3-11,-0.5) -- (3.5-11,0.866-0.5) node[above] {$v\bm{u}$};
\draw (3.5-11,-0.5) arc (0:60:0.5);
\draw[dashed] (3-11,-0.5) -- (4-11,-0.5) node[above] {$\theta$};
\draw[<->] (-5,1.1) -- node[right] {$w$} (-5,-3.1);

\path (2.5,0) node {\includegraphics[totalheight=4.25cm]{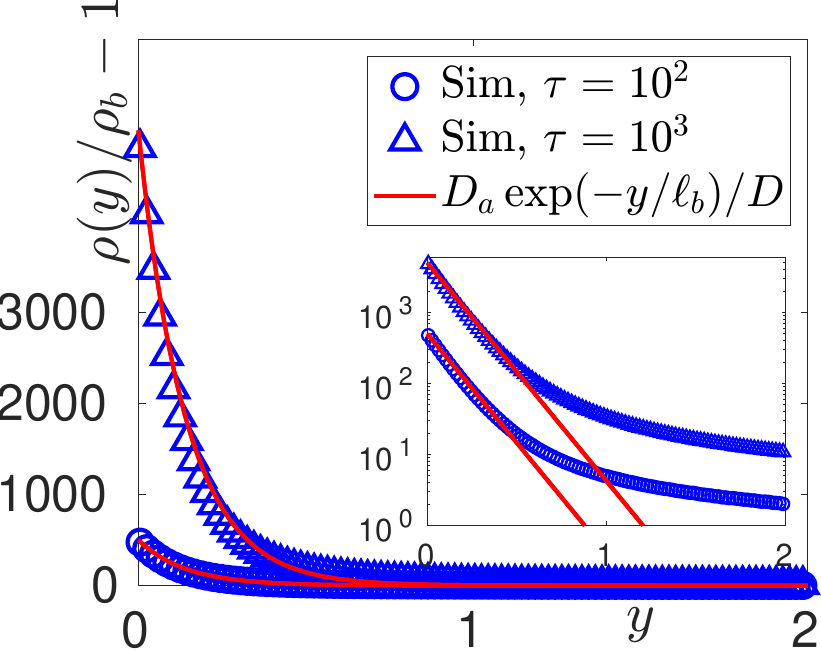}};
\node at (-2.5,2.3) {\bf b)};
\end{tikzpicture}
\end{center}
\caption{\label{fig_homo_profile} \textbf{The steady-state density profile of active Brownian particles inside tubes of constant width.} \textbf{(a)}: An illustration of an ABP moving in a homogeneous tube of width $w$. We impose periodic boundary condition in the $x$ direction. The walls are located at $y=0$ and $y=w$. \textbf{(b)}: The density profile of particles near the wall localized at $y=0$, normalized by the bulk density $\rho_b$, for different values of $\tau$. Symbols represent simulation data and solid lines represent the theoretical prediction given by Eq.~\eqref{eqn_density_profile_active_inftyw}. The inset shows the same data using a log scale for the $y$-axis. Parameters: $w=10^4$, $v=1$, $D=0.1$.}
\end{figure}

To make this article self-content, we first briefly review the
calculations of the steady-state density and polarization profiles of
ABPs in a channel with no-flux flat boundaries presented in
Ref.~\cite{Duzgun_2018}.  (For AOUPs, one can refer to
Ref.~\cite{Caprini_2018}.) The solutions will serve as ansatz of
density and polarization profiles in channels of varying widths. Other
references that assume a vanishing thermal translational
noise~\citep{Angelani_2017, Wagner_2017, Wagner_2022} are beyond the scope of this
article. The channel is invariant by translation along the $x$-axis
and $w$ is constant. The boundaries are located at $y=0$ and $w$
(Fig.~\ref{fig_homo_profile}a).

In this section, we characterize the steady-state density and
polarization across the channel. We introduce the single-particle
probability density $\rho(\rr)\equiv\langle\delta(\rr-\rr(t))\rangle$,
the average orientation field
$\mm(\rr)\equiv\langle\uu(t)\delta(\rr-\rr(t))\rangle$, and the
average nematic tensor field
$\QQ(\rr)\equiv\langle(\uu(t)\uu(t)-\bm{\mathrm{I}}/2)\delta(\rr-\rr(t))\rangle$,
where $\langle\cdot\rangle$ denotes ensemble averages with respect to
$\Phi$. We note that these fields are single-particle
observables. For $N$ non-interacting particles, replacing $\rr(t),
\uu(t)$ by $\rr_i(t),\uu_i(t)$ and summing over $i$ would lead to the
standard number-density, orientation, and nematic-order fields. The
dynamics of these fields are given by~\cite{Cates_2013, Solon_2015},
\begin{align}
\partial_t\rho=&-v\nabla\cdot \mm+D\Delta\rho\;, \\ 
\partial_t \mm=&-(v/2)\nabla \rho-v\nabla\cdot \QQ-\mm/\tau+D\Delta \mm\;.
\end{align}
Note that the dynamic equations have the same form for RTPs with
tumbling rate $\tau^{-1}$~\cite{Cates_2013, Solon_2015}.

In homogeneous channels with translational symmetry along $x$, the steady-state profiles $\rho$, $\mm$, and $\QQ$ are only functions of $y$. The boundary conditions now read 
\begin{align}
    &vm_y=D\rho'(y)\;, &v\left(\rho/2+Q_{yy}\right)=Dm_y'
\end{align}
at $y=0$ and $w$. In the steady state, the translational symmetry of the system along $x$ implies that $m_x=0$ and $Q_{yx}=0$. To close the equations, we assume $\QQ=\bm{0}$. We note that such a truncation was argued to be valid for a P\'{e}clet number $\mathrm{Pe}\equiv v\ell_p/D<100$~\cite{Row_2020}. Here, we find that this approximation leads to satisfying predictions for the number of particles accumulated near boundaries even for ${\rm Pe}\sim 10^4$ (see Fig.~\ref{fig_homo_profile}b).

In the steady state, the ABPs accumulate near the confining walls at $y=0$ and $y=w$, forming a boundary layer with a characteristic thickness 
\begin{equation}
    \ell_b\equiv D/\sqrt{(D_a+D)/\tau}\;, 
\end{equation}
where $D_a\equiv v^2\tau/2$ is the active contribution of the effective diffusion constant. If $w\gg \ell_b$, the two boundary layers are far apart, and the density profile near the homogeneous wall at $y=0$ reads
\begin{equation}
\rho_{\rm hm}(y)\simeq\rho_b\left(\frac{D_a}{D}e^{-y/\ell_b}+1\right)\;, \label{eqn_density_profile_active_inftyw}
\end{equation}
where $\rho_b$ is the bulk density at $y=w/2$. Then, the polarization profile satisfies
\begin{equation}
m_{\rm hm}(y)\equiv m_y(y)=\frac{D\rho_{\rm hm}'(y)}{v}=-\frac{D_a\rho_b}{v\ell_b}e^{-y/\ell_b}\;. \label{eqn_m_profile_active_inftyw}
\end{equation}
This result has been obtained in Refs.~\cite{Elgeti_2013, Duzgun_2018}. The density of `excess' particles (per unit length) accumulated near the boundary can then be estimated as {$\rho_b \int_0^\infty dy\, D_a e^{-y/\ell_b}/D=\rho_b D_a\ell_b/D$}.

Hence, truncating at the second order in the moment expansion of
$\Phi$, the density profile $\rho_{\rm hm}(y)$ near a hard boundary
decreases exponentially towards the bulk density $\rho_b$ over a
characteristic length $\ell_b$. We test the approximation in
agent-based simulations in Fig.~\ref{fig_homo_profile}b. Because of
translational symmetry along the $x$ direction, we only simulate the
$y$-component of Eq.~\eqref{eqn_ABP_ito} using Euler time-stepping. To
implement the no-flux boundary condition, we follow
Ref.~\cite{Duzgun_2018} and use a potential-free algorithm: If a
particle moves out of the confined region after a time step, we place
it back at its position at the beginning of the time step, while still
allowing its orientation to evolve. The numerical results show a
slightly slower decay in the bulk than the one predicted by
Eq.~\eqref{eqn_density_profile_active_inftyw}, which is due to a
nonzero value of $\nabla\cdot\QQ$. A better estimate can be obtained
by truncating at a higher-order in the moment expansion of $\Phi$,
which allows identifying more exponential components to account for a
slower decay. Overall, for our purpose with $w\gg \ell_b$,
Eq.~\eqref{eqn_density_profile_active_inftyw} provides a reasonable
estimation of $\rho_{\rm hm}(y)$, even when the persistent length of
the particle $\ell_p\equiv v\tau$ is comparable to $w$.

\section{The Fick-Jacobs framework in the limit of tubes with slow-varying width}\label{sec_FJ_active_0th}

\subsection{The generalized entropy potential}

We now turn to non-interacting ABPs/RTPs in a long channel of length
$L$ in 2d with a varying width. First, we consider the exact dynamics
of the 1d marginal density field $\tilde{\rho}(x)\equiv\int \rho(x,y)
dy$ and of the corresponding marginal for the orientation field along
the tube $\tilde{m}_1(x)\equiv\int m_x(x,y)dy$:
\begin{align}
\partial_t \tilde{\rho}=&-v\partial_x\tilde{m}_1+D\partial_{xx}\tilde{\rho} \nonumber \\
&-D\partial_x[\rho(x,w_2)w_2'(x)-\rho(x,w_1)w_1'(x)]\;, \label{eqn_tilde_rho}\\
\partial_t \tilde{m}_1=&-\frac{v}{2}\partial_x(\tilde{\rho}+\tilde{m}_2)-\frac{\tilde{m}_1}{\tau}+D\partial_{xx}\tilde{m}_1 \nonumber \\
&-D\partial_x[m_x(x,w_2)w_2'(x)-m_x(x,w_1)w_1'(x)]\;, \label{eqn_tilde_m1}
\end{align}
where $\rho(x,y)$ and $m_x(x,y)$ are defined as in
Sec.~\ref{sec_ss_homo}, and we have introduced the second moment
$\tilde{m}_2\equiv 2\int Q_{xx}dy$. Next, in order to close
Eqs~\eqref{eqn_tilde_rho}-\eqref{eqn_tilde_m1}, we employ an
approximation to establish relations between the boundary and bulk
values.

\begin{figure}
  \begin{tikzpicture}[line width=0.3mm, >=latex]
\draw[domain=-6:-2,smooth,variable=\x] plot ({\x},{0.5*(1-cos((\x-2)*360/4))+0.05});
\draw[domain=-6:-2,smooth,variable=\x] plot ({\x},{-0.5*(1-cos((\x-2)*360/4))-0.05});
\filldraw[fill=black] (1.5-6,-0.25) circle (0.015);
\draw[->] (1.5-6,-0.25) -- (1.75-6,0.433-0.25) node[above] {$v\uu$};
\draw (1.75-6,-0.25) arc (0:60:0.25);
\draw[dashed] (1.5-6,-0.25) -- (2-6,-0.25) node[above] {$\theta$};
\node at (2-6,1.25) {$y=w_2(x)$};
\node at (2-6,-1.25) {$y=w_1(x)$};
\draw[->] (0-6,-1) -- (0-6,-0.5) node[right]{$y$};
\draw[->] (0-6,-1) -- (0.5-6,-1) node[right]{$x$};

\draw (-6,1.7) node[anchor=south west] {\bf a)};
    
    \path (0.3,0) node {\includegraphics[totalheight=3.7cm]{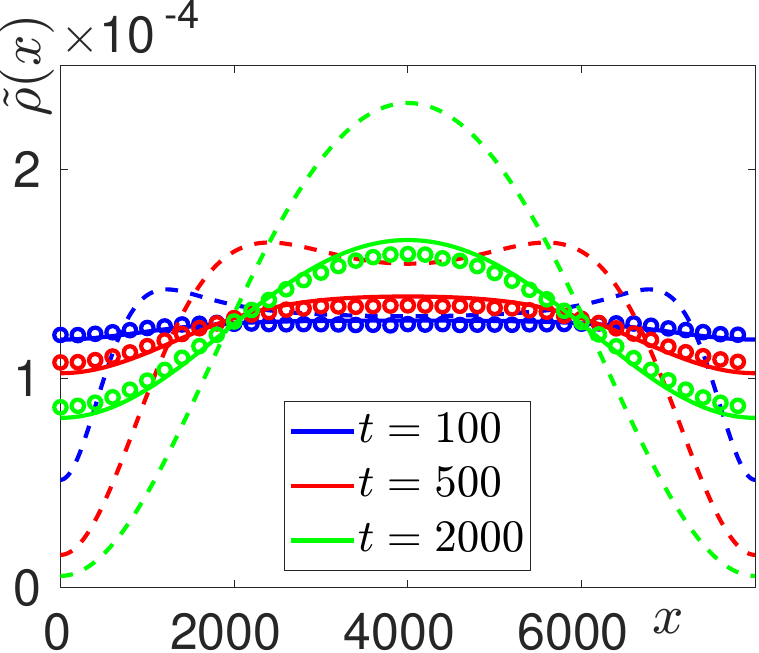}};
    \draw (-1.3,1.8) node[anchor=south west] {\bf b)};

    % \path (-4,-3) node {\includegraphics[totalheight=2.75cm]{density_inhomo_Dr0.01_Ax500.pdf}};
    % \draw (-6,1.3-3) node[anchor=south west] {\bf e)};
    
    %\path (-4,-4) node {\includegraphics[totalheight=3.4cm]{density_inhomo_changing_Dr_D.pdf}};
    %\draw (-6,1.7-4) node[anchor=south west] {\bf c)};
    
    % \path (4,-3) node {\includegraphics[totalheight=2.75cm]{density_inhomo_dual_asym_Dr0.01_Ax50_N2.pdf}};
    %\path (0.5,-4) node {\includegraphics[totalheight=3.4cm]{density_inhomo_dual_asym_Dr0.01_Ax50_N5.pdf}};
    % \draw (2,1.3-3) node[anchor=south west] {\bf g)};
    %\draw (-1.5,1.7-4) node[anchor=south west] {\bf d)};
  \end{tikzpicture}
\caption{\label{fig_inhomo_dynamics} \textbf{Dynamics of the 1d
    density profile $\tilde{\rho}(x,t)$ of ABPs along a corrugated 2d
    channel.} \textbf{(a)}: Illustration of an active particle moving
  in a spindle channel. The boundaries of the channel are located at
  $w_{1,2}(x)=\pm[A(1-\cos(2\pi x/L))/2+C_0]$. \textbf{(b)}: Time
  evolution of $\tilde{\rho}$ in a symmetric channel with $A=500$ and
  $L=8000$. Circles represent data from agent-based simulations of
  $5.12\times 10^7$ particles. The data are smoothed by averaging in a
  region of 100 length units. Solid lines shows the numerical
  solutions of Eqs.~\eqref{eqn_tilde_rho_0th_trunc} and
 ~\eqref{eqn_tilde_m1_0th_trunc} and reveal a good agreement with
  particle-based simulations. Dashed lines represent the numerical
  solution of Eq.~\eqref{eqn_FJ_passive} with a passive entropy
  potential, Eq.~\eqref{eqn_passive_potential}, which shows that the
  FJ theory cannot be straightforwardly extended to active particles
  by replacing the thermal diffusivity by an effective
  temperature. Color encodes time. Parameters: $C_0=10$, $v=5$,
  $\tau=100$, $D=0.1$.}
\end{figure}

We assume that the width of the tube changes slowly, so that $|w'(x)|
\ll 1$, leading to width variations over the tube length $L$ that are
much smaller that $L$. This allows us to assume that the channel is
locally flat and parallel to the $x$-axis, so that particles at
position $x$ obey the same local PDF $\Phi(x,y,\uu)$ as $\Phi_{\rm
  hm}(y,\uu)$ in a tube of constant width $w_{\rm hm}=w(x)$
(Fig.~\ref{fig_homo_profile}a). If $w_{\rm hm}\gg\ell_b$, the two
boundary layers are far away from each other, and they effectively
decouple so that the local density profile near each boundary wall is
given by Eq.~\eqref{eqn_density_profile_active_inftyw}. Then
\begin{equation}
    \rho(x,w_{1,2}(x))\simeq\rho_{\rm hm}(0)=\rho_b\left(\frac{D_a}{D}+1\right)\;. \label{eqn_ABP_boundary_rho}
\end{equation}
The integrated density $\tilde{\rho}$ is then the sum of a bulk
contribution, $\rho_b w(x)$, and of contributions due to the two
boundary layers, $\int_{0}^\infty [\rho_{\rm hm}(y)-\rho_b]dy$. Direct
calculations lead to
\begin{equation}
\tilde{\rho}\simeq\rho_b(w(x)+2D_a\ell_b/D)\;.  \label{eqn_ABP_int_rho}
\end{equation}
Eliminating $\rho_b$ from Eqs.~\eqref{eqn_ABP_boundary_rho} and~\eqref{eqn_ABP_int_rho} then yields
\begin{equation}
\rho(x,w_{1,2}(x))\simeq \frac{D_{\rm eff}\tilde{\rho}}{Dw(x)+2D_a\ell_b}\;, \label{eqn_ABP_boundary_rho}
\end{equation}
where $D_{\rm eff}=D_a+D$ is the effective diffusion constant of ABPs/RTPs. Substituting Eq.~\eqref{eqn_ABP_boundary_rho} into the second line of Eq.~\eqref{eqn_tilde_rho} finally leads to
\begin{align*}
&-D\partial_x[\rho(x,w_2)w_2'(x)-\rho(x,w_1)w_1'(x)] \nonumber \\
\simeq&-D\partial_x\left[\frac{D_{\rm eff}w'(x)}{Dw(x)+2D_a\ell_b}\tilde{\rho}\right]=\partial_x\left[\tilde{\rho}V'(x)\right]\;,
\end{align*}
where we have introduced the effective potential
\begin{equation}
V(x)\equiv -D_{\rm eff}\log[2D_a\ell_b/D+w(x)]\;.\label{eqn_entropy_potential_inftyw}
\end{equation}
The boundary-value problem in the dynamics of $\tilde{\rho}$ is then solved, and its dynamics is given by
\begin{equation}
\partial_t \tilde{\rho}=-v\partial_x\tilde{m}_1+D\partial_{xx}\tilde{\rho}+\partial_x\left[\tilde{\rho}V'(x)\right]\;. \label{eqn_tilde_rho_0th}
\end{equation}

Under the assumption of slow-varying width, the boundary
values $m_x(x,w_{1,2}(x))=0$ because $m_x=0$ in channels of constant
width, and the second line in Eq.~\eqref{eqn_tilde_m1} vanishes. The
boundary-value problem in the dynamics of $\tilde{m}_1$ is
automatically solved without the emergence of an effective potential,
\begin{equation}
\partial_t \tilde{m}_1=-(v/2)\partial_x(\tilde{\rho}+\tilde{m}_2)-\tilde{m}_1/\tau+D\partial_{xx}\tilde{m}_1\;,\label{eqn_tilde_m1_0th}
\end{equation}
so that the dynamics of $m_1$ is independent of the confinement. In
line with the truncation $\QQ=0$ applied to obtain
Eqs.~\eqref{eqn_density_profile_active_inftyw} and
\eqref{eqn_m_profile_active_inftyw}, we truncate the moment expansion
by assuming $\tilde{m}_2=0$ to close Eqs.~\eqref{eqn_tilde_rho_0th}
and~\eqref{eqn_tilde_m1_0th}. From now on and consider the following
dynamics:
\begin{align}
\partial_t \tilde{\rho}=&-v\partial_x\tilde{m}_1+D\partial_{xx}\tilde{\rho}+\partial_x\left[\tilde{\rho}V'(x)\right]\;, \label{eqn_tilde_rho_0th_trunc} \\
\partial_t \tilde{m}_1=&-(v/2)\partial_x\tilde{\rho}-\tilde{m}_1/\tau+D\partial_{xx}\tilde{m}_1\;,\label{eqn_tilde_m1_0th_trunc}
\end{align}
which are the first central result of this article.

The closure $\tilde{m}_2=0$ can be justified by considering the
relaxation time of the moments $\tilde{m}_n=\int\Phi(x,y,\uu,t)\cos
(n\theta)d\theta dy$. Since the density field is a conserved field,
its relaxation time diverges as $L\to \infty$. On the contrary, the
relaxation time of $\tilde{m}_1$ scales as $\tau$, which is finite in
the $L\to\infty$ limit. Similarly, $\tilde{m}_n$ has a relaxation
time of the order of $\tau/n^2$~\cite{Solon_2015}. Thus, at large
times, the moments of order $n\geq 1$ are fast modes enslaved to
$\tilde{\rho}$. The dynamics of $\tilde{\rho}$ is modulated by the
change in tube width, which are assumed to be small. This allows us to
expand all observables in derivatives of $\tilde{\rho}$. Since
$\tilde{m}_n$ are of the order of $\partial_x^n\tilde{\rho}$, keeping
only the moments up to order $\partial_x\tilde{\rho}$ leads to the
truncation $\tilde{m}_2=0$ in Eqs.~\eqref{eqn_tilde_rho_0th_trunc} and
\eqref{eqn_tilde_m1_0th_trunc}.

In summary, under the assumption of a slow-varying $w(x)$, the
dynamics of the particle density along the tube is equivalent to that
of self-propelled particles experiencing a generalized entropy
potential $V(x)$ given by Eq.~\eqref{eqn_entropy_potential_inftyw},
while the polarization field evolves as if in a free space. Note that
the effective potential found for the active particles differs from
the passive one in two respects: First, the passive diffusivity $D$
entering $V_{\rm eqm}$ has to be replaced by $D_{\rm eff}$; Second,
the area $A(x)$ is replaced by $w(x)+ 2D_a\ell_b/D$. This is a
signature of the accumulation of active particles near
boundaries---the number of excess particles accumulated near the two
boundaries is $2n_{\rm bd}\equiv 2D_a\ell_b/D$. This `buffer' of
active particles lower the impact of the variations of the tube width
on the dynamics of $\rho(x,t)$. This can be understood as
follows. Since the width of the boundary layer scales as $\ell_b\sim
D_a^{-1/2}$ at large $D_a$, we find that $n_{\rm bd}\sim
D_a^{1/2}$. In this limit, a thin boundary layer thus contributes
significantly to the total density. The particles then spend most of
the time near the boundaries. The number of particles absorbed by the boundary is independent of the width of the 2d channel, so that the particles have less time to sample the change
in the channel width.

In Fig.~\ref{fig_inhomo_dynamics}b, we compare the dynamics given by
Eqs.~\eqref{eqn_tilde_rho_0th_trunc} and
\eqref{eqn_tilde_m1_0th_trunc} with the result of agent-based
simulations. Starting from an initially density such that
$\tilde{\rho}(x,0)$ is a constant, we simulate ABPs in a spindle
channel whose width varies according to $w(x)=A(1-\cos(2\pi
x/L))+2C_0$ and impose no-flux boundary conditions. The initial
position $y$ and orientation $\theta$ of the particle are random. We
then measure $\tilde{\rho}(x,t)$ in simulations (circles) at different
times, and compare it with the numerical solutions (solid curves) of
Eqs.~\eqref{eqn_tilde_rho_0th_trunc}
and~\eqref{eqn_tilde_m1_0th_trunc}. We use a semi-spectral method and
4th order Adams–Bashforth time-stepping to solve the partial
differential equations. To highlight the importance of the boundary
layer to $V(x)$, we compare our simulations with those of a fully
passive dynamics Eq.~\eqref{eqn_FJ_passive} with $D$ replaced by
$D_{\rm eff}$ (dashed curves). Clearly, our generalized potential
Eq.~\eqref{eqn_entropy_potential_inftyw} captures better the dynamics
of ABPs.

Note that, in this section, we truncated the moment expansion by
setting $m_{n\geq 2}=0$. In principle, we can continue the moment
expansion and calculate the dynamics of $\tilde{m}_2$. We expect that
a boundary term of the form
$2D\partial_x[Q_{xx}(x,w_2)w'_2(x)-Q_{xx}(x,w_1)w'_1(x)]$ will appear
in the evolution equation, yielding an entropy potential for
$\tilde{m}_2$ (that will be generically distinct from that entering
the dynamics of $\rho$). To continue this procedure for higher order
moments $\tilde{m}_n$ with $n>2$ would require computing the
steady-state moments in channels of constant width, which is beyond
the scope of this article.

\subsection{Steady-state density profiles  in the diffusive limit}\label{sec_ss_boltzmann}

\begin{figure}
  \begin{tikzpicture}[line width=0.3mm, >=latex]
    \path (-4,4) node {\includegraphics[totalheight=3.4cm]{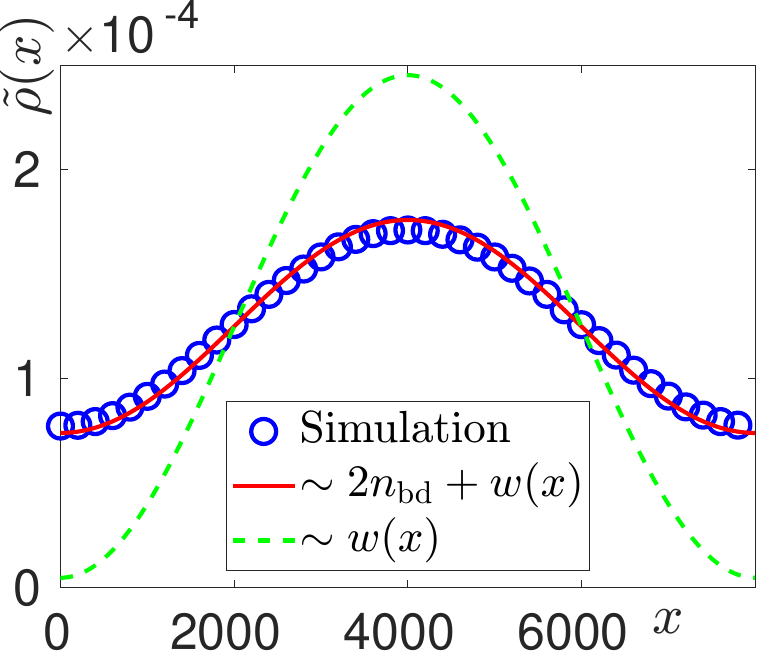}};
    \draw (-6,1.7+4) node[anchor=south west] {\bf a)};
    
    \path (0.5,+4) node {\includegraphics[totalheight=3.4cm]{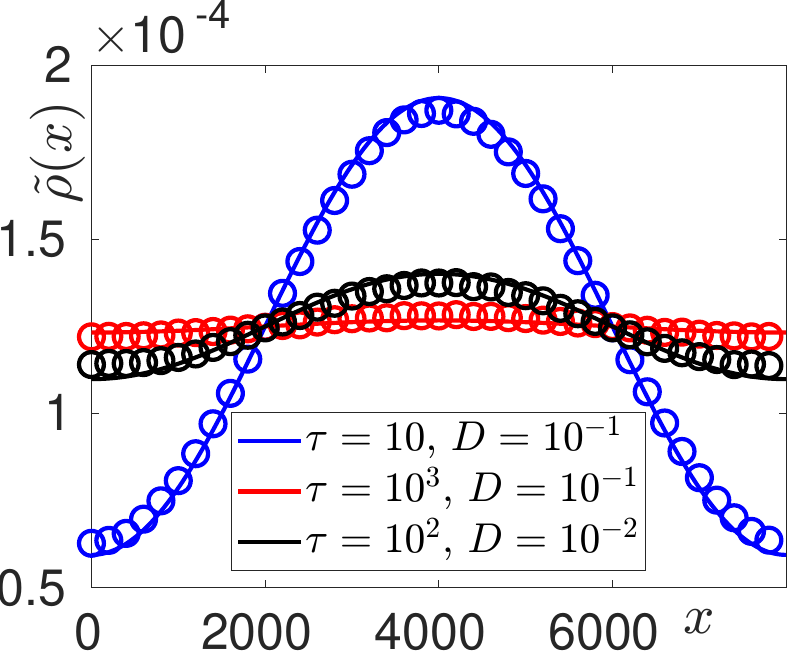}};
    \draw (-1.5,1.7+4) node[anchor=south west] {\bf b)};
    
\draw[domain=-6:-2,smooth,variable=\x] plot ({\x},{0.5*(1-cos(((\x+6)/4)^5*360))+0.05});
\draw[domain=-6:-2,smooth,variable=\x] plot ({\x},{-0.5*(1-cos(((\x+6)/4)^5*360))-0.05});
\filldraw[fill=black] (1.5-4,-0.25) circle (0.015);
\draw[->] (1.5-4,-0.25) -- (1.75-4,0.433-0.25);

\draw (-6,1.7) node[anchor=south west] {\bf c)};
    
    \path (0.5,0) node {\includegraphics[totalheight=3.4cm]{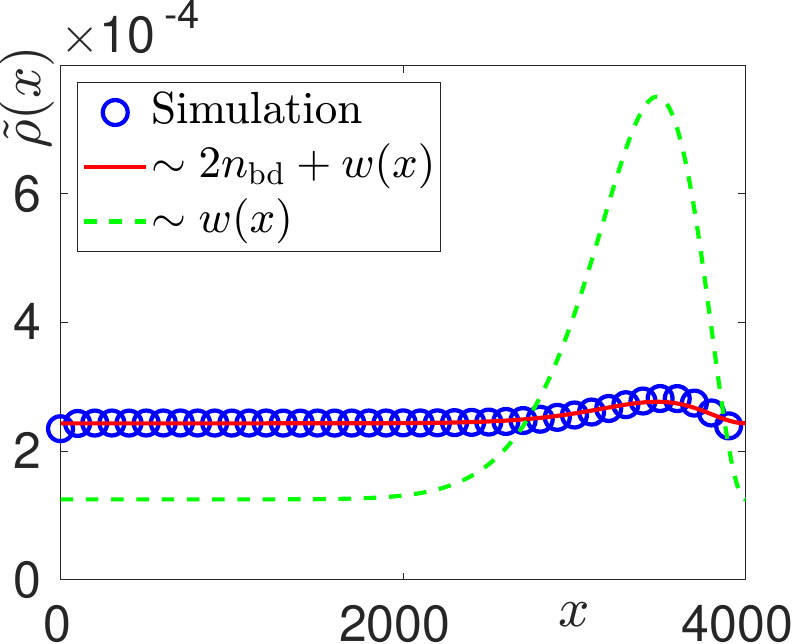}};
    \draw (-1.5,1.7) node[anchor=south west] {\bf d)};
    
  \end{tikzpicture}
\caption{\label{fig_inhomo_profile} \textbf{Steady-state density
    profiles of ABPs along a corrugated 2d channel.} \textbf{(a,b)}:
  Density profiles in symmetric channels with
  $w_{1,2}(x)=\pm[A(1-\cos(2\pi x/L))/2+C_0]$. \textbf{(c,d)}:
  Asymmetric channel with $w_{1,2}=\pm[A(1-\cos(2\pi (x/L)^5))/2+C_0]$
  \textbf{(c)} and the corresponding density profile of ABPs
  \textbf{(d)}. In all panels, circles represent the simulation
  data. Solid lines are the theoretical estimation obtained by
  Eq.~\eqref{eqn_general_boltzmann}. Dashed lines are obtained using
  the passive counterpart Eq.~\eqref{eqn_passive_potential}. In
  \textbf{(a)}, we choose $A=500$ and $L=8000$. In \textbf{(b)}, we
  show that our theoretical predictions hold for a large range of
  values of $\tau$ and $D$. $A=100$ and $L=8000$. In \textbf{(d)}, we
  choose $A=50$ and $L=4000$. If not specified, we take $C_0=10$,
  $v=5$, $\tau=100$, $D=0.1$ in the simulations.}
\end{figure}

We now consider the steady state of
Eqs.~\eqref{eqn_tilde_rho_0th_trunc} and
\eqref{eqn_tilde_m1_0th_trunc}. The steady state of
Eq.~\eqref{eqn_tilde_m1_0th_trunc} imposes
\begin{equation*}
\tilde{m}_1=-(v\tau/2)\partial_x\tilde{\rho}+D\tau\partial_{xx}\tilde{m}_1\;.
\end{equation*}
Substituting this into Eq.~\eqref{eqn_tilde_rho_0th_trunc} then leads
to
\begin{equation*}
\left(\frac{v^2\tau}{2}+D\right)\partial_{xx}\tilde{\rho}+\partial_x\left[\tilde{\rho}V'(x)\right]-vD\tau\partial_{x}^3\tilde{m}_1=0\;. \end{equation*}
Since $\partial_{x}^3\tilde{m}_1\sim\partial_x^4\tilde{\rho}$, we neglect this term in the limit of large $L$ and slow-varying $w(x)$. In periodic tubes, $w(0)=w(L)$ and the  density of active particles along the tube has the form of a Boltzmann weight with an effective temperature $D_{\rm eff}$, 
\begin{equation}
\tilde{\rho}_{\rm B}(x)\propto \exp(-V(x)/D_{\rm eff})=2n_{\rm bd}+w(x)\;. \label{eqn_general_boltzmann}
\end{equation} 
This prediction can be compared by agent-based simulations with both
channels of symmetric (Fig.~\ref{fig_inhomo_profile}a,b) and
asymmetric (Fig.~\ref{fig_inhomo_profile}c,d) shapes. Compared to the
passive predictions, the 1d density profiles are flattened by the
existence of boundary layers. We note that the
approximation does not hold at position where $|w'(x)|$ is not small.

\subsection{Generalization to 3d tubes}\label{sec_3d}

Our construction of the effective potential can also be applied to the
case of 3d tubes. We consider tubes along the $x$ direction without
loss of generality. We note that a large curvature of the boundary in
the $yz$ plane can also induce particle accumulations, and the full
solution of the density profile near the curved wall can be
nontrivial~\cite{Nikola_2016}. Suppose the radius of the boundaries of
cross-sections is larger than $\ell_p$ such that the boundary is
locally flat. Then the density profile near the boundary can be
approximated again by
Eq.~\eqref{eqn_density_profile_active_inftyw}. The integration of the
density profile in a cross-section in the $yz$ plane can be estimated
as the sum of contributions from bulk and the boundary. We have
$\tilde{\rho}(x)\approx\rho_b [A(x)+n_{\rm bd}P(x)]$, where $A(x)$ is
the area of the cross-section of the tube along the $yz$ plane at $x$,
and $P(x)$ is the perimeter of the cross-section. Following the same
procedures as in Sec.~\ref{sec_FJ_active_0th}, we have
\begin{equation}
V(x)=-D_{\rm eff}\log\left[n_{\rm bd}P(x)+A(x)\right]\;.\label{eqn_entropy_potential_3D}
\end{equation} 
Note that $n_{\rm bd}=D_a\ell_b/D$ and $D_a=v^2\tau/d$ in $d$-dimensional space.

\begin{figure}
  \begin{tikzpicture}
    \path (-4.5,0) node {\includegraphics[totalheight=3.4cm]{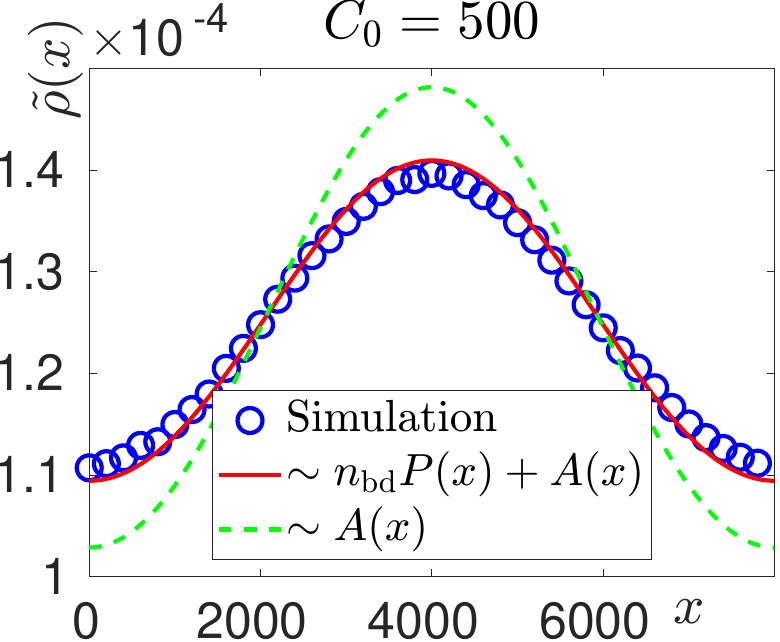}};
    \path (0,0) node {\includegraphics[totalheight=3.4cm]{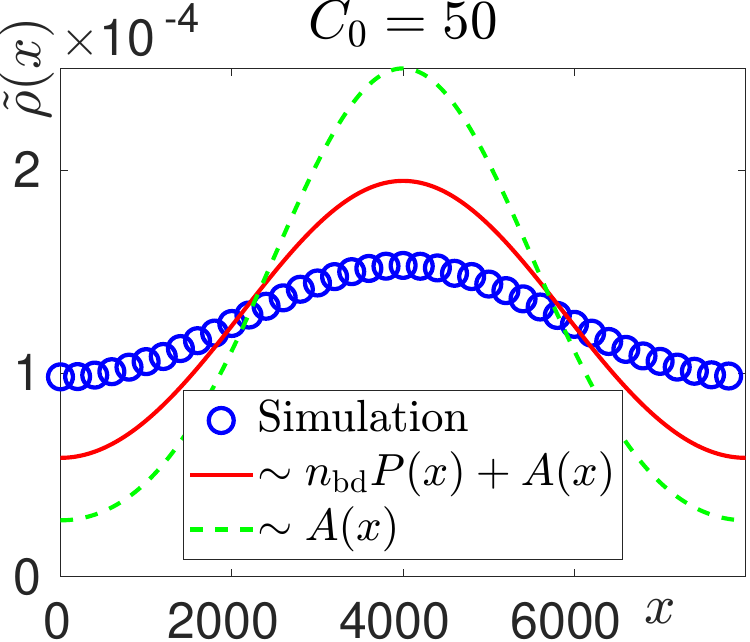}};
    \draw (-6.5,1.7) node[anchor=south west] {\bf a)};
    \draw (-2,1.7) node[anchor=south west] {\bf b)};
  \end{tikzpicture}
\caption{\label{fig_inhomo_profile_3D} \textbf{The 1d density profile of RTPs along a corrugated 3d tube.} The cross sections in $yz$ plane is circular of radius $R(x)=R_0[1-\cos(2\pi x/L)]/2+C_0$. $C_0=500$ \textbf{(a)} and $50$ \textbf{(b)}. Circles represent data from agent-based simulations. The theoretical estimation of the density profile is obtained by $\exp(-V(x)/D_{\rm eff})$. The solid lines are calculated using the $V(x)$ in Eq.~\eqref{eqn_entropy_potential_3D}, while the dashed lines are using the passive counterpart Eq.~\eqref{eqn_passive_potential}. $L=8\times 10^3$, $R_0=100$, $v=5$, $\tau=100$, $D=0.1$.}
\end{figure}

The steady-state marginal density along the tube can then be estimated
from Eq.~\eqref{eqn_entropy_potential_3D} and compared with
simulations of RTPs in a 3d tube with rotational symmetry along the
$x$ axis (Fig.~\ref{fig_inhomo_profile_3D}a). Note that the estimation
is valid only for sufficiently large $C_0$ to ensure that the
curvature of the tube is negligible. Otherwise, the accumulation of
particles along a curved boundary will further flatten the density
profile (Fig.~\ref{fig_inhomo_profile_3D}b). The agreement in
Fig.~\ref{fig_inhomo_profile_3D}a also indicates that our previous
analysis, detailed for ABPs, directly applies to RTPs. This is
expected since ABPs and RTPs with the same values of $v_0$ and $\tau$
share the same dynamics for $\rho$ and $\mm$.

\subsection{Mean-escape time of an active particle from a spindle chamber} \label{sec_mfpt}

With the generalized entropy potential
\eqref{eqn_entropy_potential_inftyw}, one can calculate the
mean-first-passage time (MFPT) of one particle escaping from a spindle
chamber (Fig.~\ref{fig_inhomo_profile}a). The mean-first-passage time
$t_{\rm MFP}$ of a particle starting from the middle of the spindle
$x=L/2$ with random orientation and reaching one of the necks at $x=0$
or $L$ is given by Kramer's seminal work~\cite{Kramers_1940,
  note_oneforth}
\begin{equation}
t_{\rm MFP}\sim \frac{\pi}{2\sqrt{|V''(L/2)V''(0)|}}\exp\frac{V(0)-V(L/2)}{D_{\rm eff}}\;.\label{eqn_mfpt}
\end{equation}
For passive particles, substituting Eq.~\eqref{eqn_passive_potential} leads to $t_{\rm MFP}\propto  1/D$.

\begin{figure}
\centering
  \begin{tikzpicture}[line width=0.3mm, >=latex]
    \path (-4.5,0) node {\includegraphics[totalheight=3.4cm]{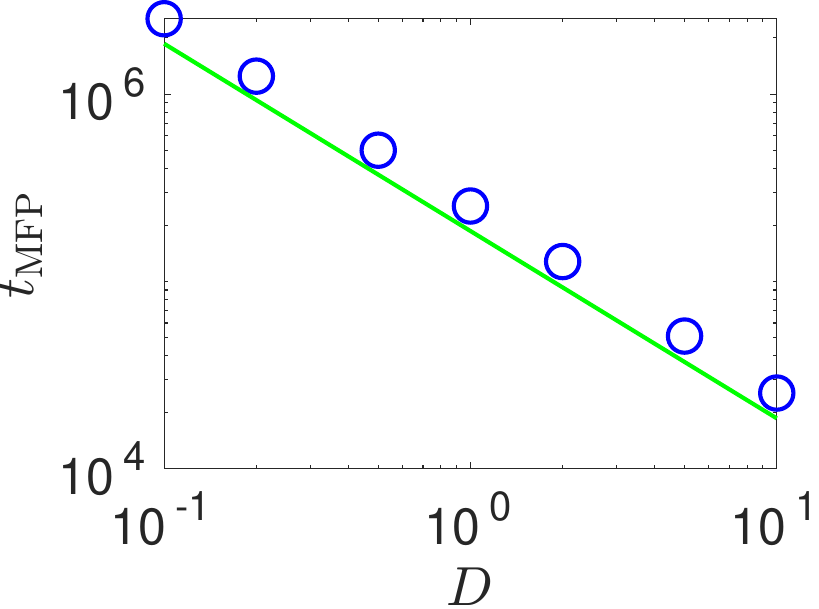}};
    \path (0,0) node {\includegraphics[totalheight=3.4cm]{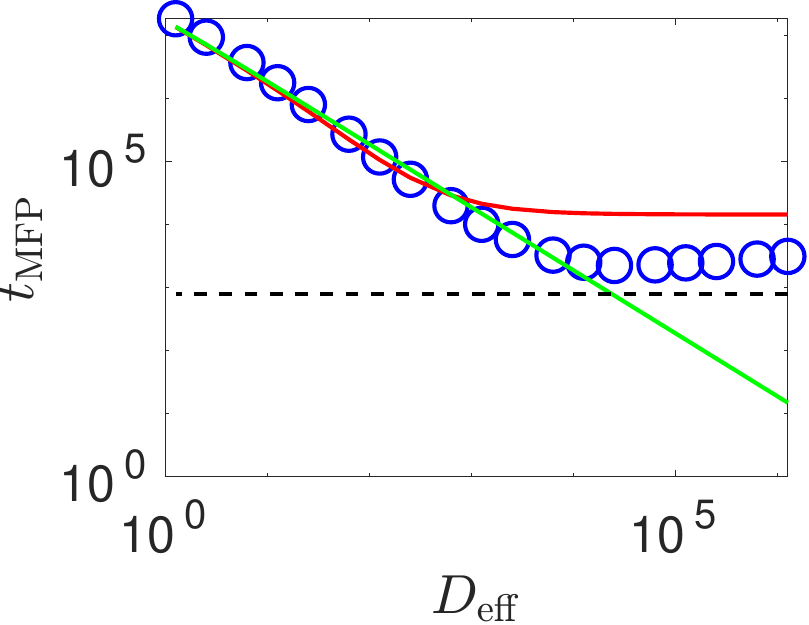}};
    \draw (-6.5,1.7) node[anchor=south west] {\bf a)};
    \draw (-2,1.7) node[anchor=south west] {\bf b)};

    \draw[thin] (-5.85+0.1,-0.1) rectangle (-3.75+0.1,-0.7);
    \draw[domain=-5.8+0.1:-3.8,smooth,variable=\x] plot ({\x},{(1-cos((\x+5.8-0.1)*360/2))*0.1+0.05-0.4});
    \draw[domain=-5.8+0.1:-3.8,smooth,variable=\x] plot ({\x},{(-1+cos((\x+5.8-0.1)*360/2))*0.1-0.05-0.4});
    \filldraw[fill=black] (-4.8+0.1,-0.4) circle (0.015);

    \draw[thin] (-5.85+4.6,-0.1) rectangle (-3.75+4.6,-0.7);
    \draw[domain=-5.8+4.6:-3.8+4.6,smooth,variable=\x] plot ({\x},{(1-cos((\x+5.8-4.6)*360/2))*0.05+0.05-0.4});
    \draw[domain=-5.8+4.6:-3.8+4.6,smooth,variable=\x] plot ({\x},{(-1+cos((\x+5.8-4.6)*360/2))*0.05-0.05-0.4});
    \filldraw[fill=black] (-4.8+4.6,-0.4) circle (0.015);
    \draw[->] (-4.8+4.6,-0.4) -- (-4.8+0.25*0.5+4.6,0.433*0.5-0.4);
    
    \path (-4.5,-3.8) node {\includegraphics[totalheight=3.4cm]{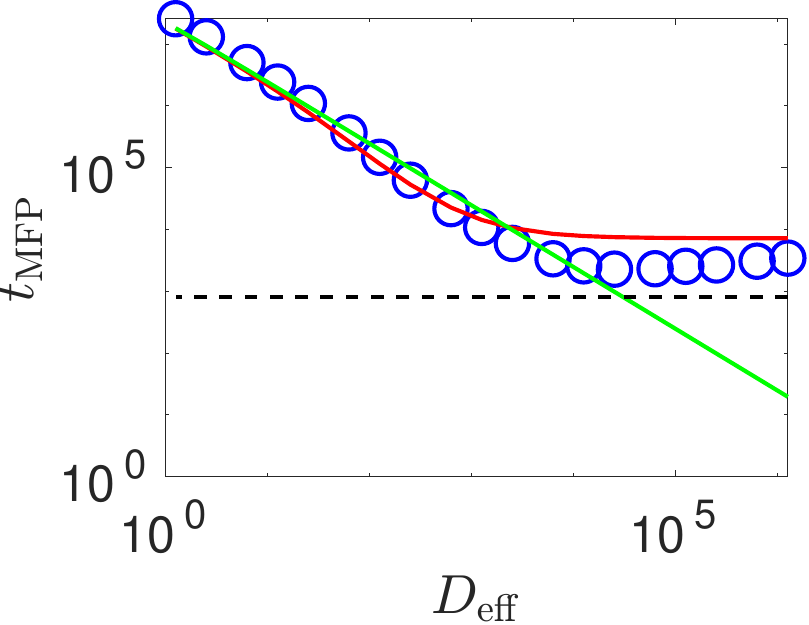}};
    \path (0,-3.8) node {\includegraphics[totalheight=3.4cm]{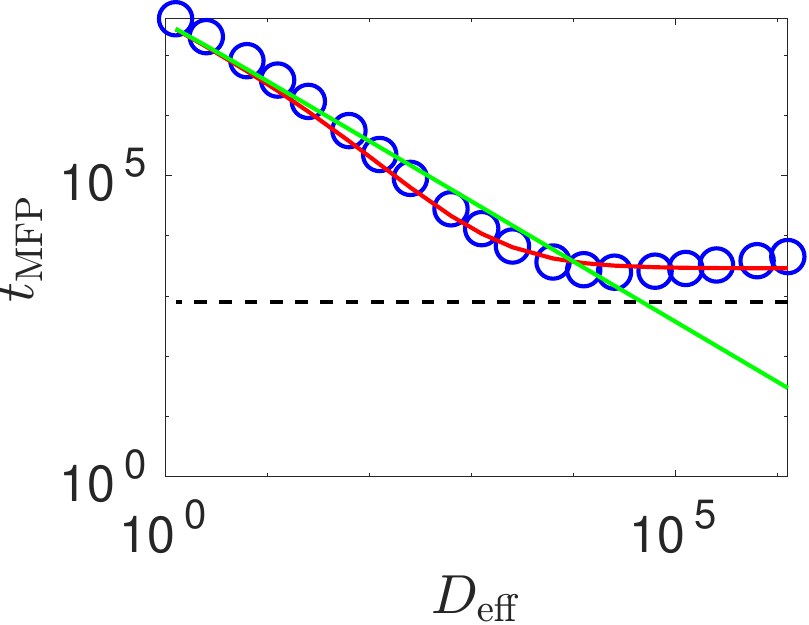}};

    \draw[thin] (-5.85+0.1,-3.8-0.1) rectangle (-3.75+0.1,-3.8-0.7);
    \draw[domain=-5.8+0.1:-3.8+0.1,smooth,variable=\x] plot ({\x},{(1-cos((\x+5.8-0.1)*360/2))*0.1+0.05-0.4-3.8});
    \draw[domain=-5.8+0.1:-3.8+0.1,smooth,variable=\x] plot ({\x},{(-1+cos((\x+5.8-0.1)*360/2))*0.1-0.05-0.4-3.8});
    \filldraw[fill=black] (-4.8+0.1,-0.4-3.8) circle (0.015);
    \draw[->] (-4.8+0.1,-0.4-3.8) -- (-4.8+0.25*0.5+0.1,0.433*0.5-0.4-3.8);
    
    \draw[thin] (-5.85+4.6,-0-3.8) rectangle (-3.75+4.6,-0.8-3.8);
    \draw[domain=-5.8+4.6:-3.8+4.6,smooth,variable=\x] plot ({\x},{(1-cos((\x+5.8-4.6)*360/2))*0.15+0.05-0.4-3.8});
    \draw[domain=-5.8+4.6:-3.8+4.6,smooth,variable=\x] plot ({\x},{(-1+cos((\x+5.8-4.6)*360/2))*0.15-0.05-0.4-3.8});
    \filldraw[fill=black] (-4.8+4.6,-0.4-3.8) circle (0.015);
    \draw[->] (-4.8+4.6,-0.4-3.8) -- (-4.8+0.25*0.5+4.6,0.433*0.5-0.4-3.8);
    
    \draw (-6.5,1.7-3.8) node[anchor=south west] {\bf c)};
    \draw (-2,1.7-3.8) node[anchor=south west] {\bf d)};
  \end{tikzpicture}
\caption{\label{fig_MFPT} \textbf{The mean first passage time of a passive particle (a) or an ABP (b-d) moving from the middle of a spindle ($x=L/2$) to one of the necks ($x=0$ or $L$).} The insets in (a) and (b) show the setup of the spindle. The width of the spindle is $w(x)=A[1-\cos(2\pi x/L)]+2C_0$. $L=8000$, $C_0=10$, $A=100$ \textbf{(b)}, 200 \textbf{(c)}, and 500 \textbf{(d)}. The simulation data are shown as circles. The red lines are estimated from the active effective potential Eq.~\eqref{eqn_entropy_potential_inftyw}, while the green lines are obtained from Eq.~\eqref{eqn_passive_potential}. The black dashed lines indicate the running time $L/(2v)$ for a particle from $x=L/2$ to one of the necks. $v=5$ and $D=0.1$ for the ABPs, and $\tau$ ranges from $10^{-1}$ to $10^5$. }
\end{figure}

The $1/D$ scaling of $t_{\rm MFP}$ for passive particles can be
checked in simulations (Fig.~\ref{fig_MFPT}a). $t_{\rm MFP}$ for
active particles, however, does not follow the $1/D_{\rm eff}$ scaling
because the activity contributes to both the effective diffusion
constant and the effective width of the channel. $t_{\rm MFP}$ first
decreases with $D_{\rm eff}$, similar to passive particles, and then
increases after an optimal $D_{\rm eff}$
(Fig.~\ref{fig_MFPT}b-d). Unlike passive particles, ABPs have finite
moving speed, and the time for ABPs to travel $L/2$ is bounded by
$L/(2v)$. For larger $D_{\rm eff}$, the time for ABPs to stay near the
boundary increases with $\tau$, which explains the increase of escape
time for ABPs from the channel. Note that Eq.~\eqref{eqn_mfpt} with
the effective potential Eq.~\eqref{eqn_entropy_potential_inftyw}
qualitatively describe the transition away from the passive regime. At
large $\tau$ the condition that $w\gg \ell_p$ cannot be held and
Eq.~\eqref{eqn_entropy_potential_inftyw} fails. Thus, quantitative
agreement in the large $D_{\rm eff}$ limit cannot be expected.

%\begin{figure}
%\centering
%  \begin{tikzpicture}
%    \path (-4,0) node {\includegraphics[totalheight=2.75cm]{mfpt_passive.pdf}};
%    \path (0,0) node {\includegraphics[totalheight=2.75cm]{mfpt_D0.1_A500_L8000.pdf}};
%    \draw (-6,1.3) node[anchor=south west] {\bf a)};
%    \draw (-2,1.3) node[anchor=south west] {\bf b)};
%    
%%    \path (-4,-3) node {\includegraphics[totalheight=2.75cm]{mfpt_D0.1_A200_L8000.pdf}};
%%    \path (0,-3) node {\includegraphics[totalheight=2.75cm]{mfpt_D0.1_A500_L8000.pdf}};
%%    \draw (-6,1.3-3) node[anchor=south west] {\bf c)};
%%    \draw (-2,1.3-3) node[anchor=south west] {\bf d)};
%  \end{tikzpicture}
%\caption{\label{fig_MFPT} \textbf{The mean first passage time of a passive particle (a) or an ABP (b) moving from the middle of a spindle ($x=L/2$) to one of the necks ($x=0$ or $L$).} The width of the spindle is defined as $w(x)=A[1-\cos(2\pi x/L)]+2C_0$. $L=8000$, $C_0=10$. $A=100$ (a) and 500 (b). The simulation data are shown by the circles. The red lines are estimated from the effective potential given by Eq.~\eqref{eqn_entropy_potential_inftyw}, while the green lines are from Eq.~\eqref{eqn_mfpt} with $V=-D_{\rm eff}\log w(x)$. The black dashed lines indicate the running time $L/(2v)$ for a particle from $x=L/2$ to $x=0$ or $L$. $v=5$ and $D=0.1$ for the ABPs, and $\tau$ ranges from $10^{-1}$ to $10^5$. }
%\end{figure}

\section{Emergence of ratchet flow: Beyond the slow-varying width approximation}\label{sec_FJ_active_1st}

\begin{figure}

\begin{tikzpicture}[line width=0.3mm, >=latex]
%\draw[domain=-6:-2,smooth,variable=\x] plot ({\x},{0.5*(1-cos(((\x+6)/4)^5*360))+0.05});
%\draw[domain=-6:-2,smooth,variable=\x] plot ({\x},{-0.5*(1-cos(((\x+6)/4)^5*360))-0.05});
%\draw[->] (1.5-4.7,-0) -- (2-4.7,-0) node[right] {$J_{\rm tot}$};

\draw (-3.5,3.8) node[anchor=south west] {\bf a)};
\path (0.5,3) node {\includegraphics[width=8.6cm]{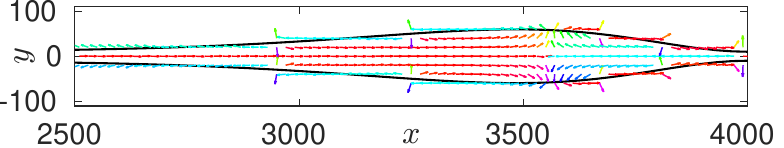}};
    
    \path (0.5,0) node {\includegraphics[totalheight=3.4cm]{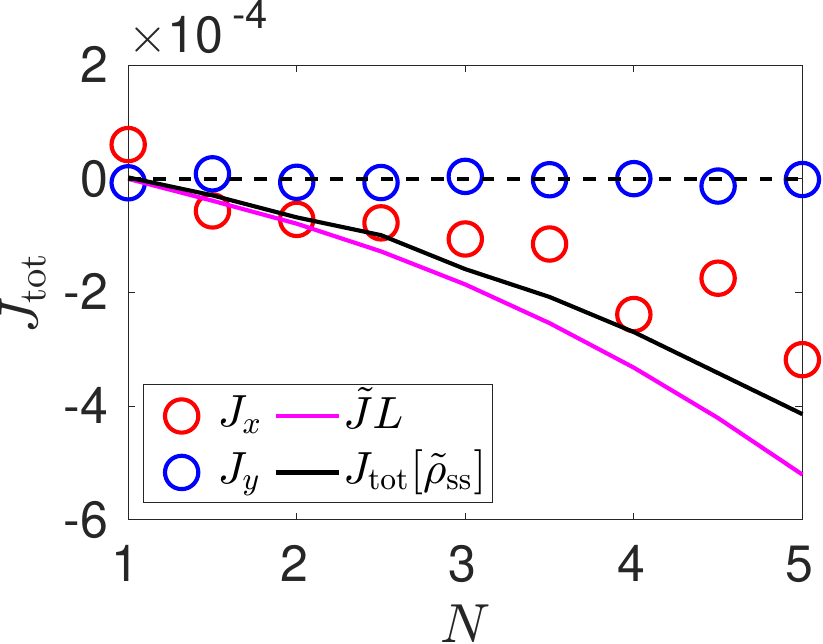}};
    \draw (-1.5,1.7) node[anchor=south west] {\bf b)};
\end{tikzpicture}

\caption{\label{fig_ratchet_flow} \textbf{The ratchet flow of ABPs in asymmetric 2d channels.} The boundary is located at $y=w_{1,2}=\pm[A(1-\cos(2\pi (x/L)^N))/2+C_0]$ with varying $N$. \textbf{(a)} The map of 2d currents $\J$ in an asymmetric channel with $N=5$. The black curves denote the boundaries. The size of arrows is proportional to $\log|\J/10^{-10}|$, and the color codes direction. \textbf{(b)} The magenta curve represents $\tilde{J}L$ calculated by Eq.~\eqref{eqn_tilde_J_sol}. The solid black curve shows the estimation of the total flow $J_{\rm tot}[\tilde{\rho}_{\rm ss}]$ obtained by Eq.~\eqref{eqn_J_tot}, where $\tilde{\rho}_{\rm ss}(x)$ is measured from the simulations. The dashed black line shows $J=0$. $A=50$, $C_0=10$, $v=5$, $\tau=100$, $D=0.1$.}
\end{figure}

The presence of spatial anisotropy in non-equilibrium systems often
lead to the emergence of spontaneous current~\cite{Magnasco_1993}, a
phenomenon referred to as ratchet current. This phenomenon has
attracted a lot of interest~\cite{Leonardo_2010,angelani2011active, Ai_2014,
  Malgaretti_2017,reichhardt2017ratchet} but there is no general
principle to predict the amplitude of ratchet currents and their
dependence on the details of the systems.  Active particles moving in
channels with broken parity symmetry constitute a setup with ratchet
currents~\cite{angelani2011active,Ai_2014, Malgaretti_2017}. We show
an example of ratchet flows of active particles in an asymmetric tube
in Fig.~\ref{fig_ratchet_flow}a.

Before detailing the underlying mechanism, we stress that ratchet
currents emerge only in systems with broken detailed balance. The
Boltzmann-like steady state given by Eq.~\eqref{eqn_general_boltzmann}
obeys detailed balance so that no macroscopic currents survive with
the approximation considered so far. This indicates that ratchet
currents do not survive in the $L\to\infty$ limit with vanishing
$w'(x)$. To observe a non-zero current, we thus have to work to higher
order. In this section we discuss how corrections can be introduced
into our formalism to account for finite $w'(x)$, and how these
corrections lead to the emergence of a ratchet flow.

\subsection{The generalized entropy potential for the polarization field}

\begin{figure}
    \centering
    \begin{tikzpicture}[line width=0.3mm, >=latex]
        
        \draw (-1.8,-0.05) -- (1.8,-0.95);
        
        \draw (-0.1,-0.475) -- (-0.1+0.025,-0.475+0.1) -- (-0.1+0.025+0.1,-0.475+0.1-0.025);
        \draw[->] (0,-0.5) -- (0+0.4,-0.5+1.6) node[above] {$\mm$};
        \draw[->] (0,-0.5) -- (0+0.4,-0.5) node[right] {$\mm_{x}$};
        \draw (0.3,-0.5) -- (0.3,-0.4) -- (0.4,-0.4);
        \draw[dashed] (0+0.4,-0.5) -- (0+0.4,-0.5+1.6);
        
        \draw[dashed] (-0.5,-0.75) -- node[below] {1} (1,-0.75);
        \draw[dashed] (-0.5,-0.375) -- node[left] {$|w_2'(x)|$} (-0.5,-0.75);
        \draw (-0.5,-0.65) -- (-0.4,-0.65) -- (-0.4,-0.75);
        
        \draw (-2,1.7) node[anchor=south west] {\bf a)};
        
        \path (4.3,0) node {\includegraphics[totalheight=3.6cm]{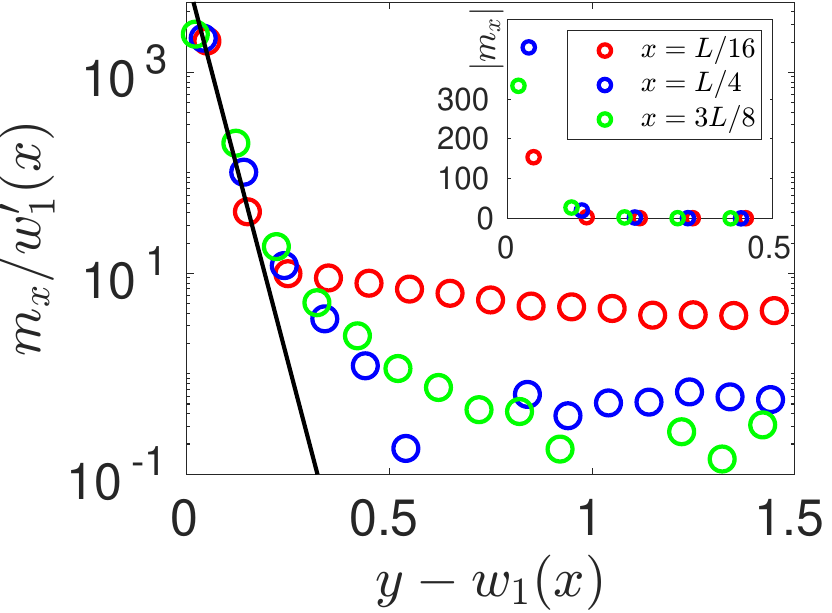}};
        \draw (2.3,1.7) node[anchor=south west] {\bf b)};

    \end{tikzpicture}
    \caption{\label{fig_corrections} \textbf{Corrections brought by the tilted flat boundary.} \textbf{(a)}: An illustration of the setup of a tilted flat boundary, where the boundary has a slope of $w_{1,2}'(x)$. We show the upper boundary for instance. \textbf{(b)}: The numerical justification of the assumption. Circles represent measured polarization profile at different $x$ from agent-based simulations, scaled by the local tangential slope of the boundary $w_1'(x)$. The solid line represents shifted profile $|m_{\rm hm}|$ given by Eq.~\eqref{eqn_m_profile_active_inftyw} with $y\to y-w_1(x)$. Inset shows the magnitude of $m_x$ in linear scale. We use a symmetric channel with $A=50$, $L=800$. }
\end{figure}

To account for a non-vanishing $w'(x)$, we assume that the wall at
position $x$ is locally flat, but tilted with a slope of $w_{1,2}'(x)$
(See Fig.~\ref{fig_corrections}a). In the limit of wide channels,
$w\gg \ell_b$, the boundary layers near the top and bottom walls
decouple and we can assume a local translational symmetry along the
tangential direction of the boundaries. Then, at the boundary, $\mm$
is locally normal to the boundary with a magnitude given by $|m_{\rm
  hm}(h)|$ in Eq.~\eqref{eqn_m_profile_active_inftyw}, where $h$ is
the distance from the wall. Locally, $m_x(x,y)$ is $\mm$ projected on
the $x$-axis. From Fig.~\ref{fig_corrections}a we have
\begin{align}
m_x(x,w_{2}(x))&=-|m_{\rm hm}(0)|\sin\arctan w_2'(x) \nonumber \\
&\simeq -|m_{\rm hm}(0)|w_2'(x)=-D_a\rho_bw_2'(x)/(v\ell_b)\;, \label{eqn_boundary_m}
\end{align}
and similarly $m_x(x,w_{1}(x))\simeq D_a\rho_bw_1'(x)/(v\ell_b)$. This approximation can be justified in Fig.~\ref{fig_corrections}b, where we plot $m_x(x,y)/w_1'(x)$ at different position $x$ of ABPs in a spindle channel in the steady state of an agent-based simulation. The simulation data shown as circles are compared with $|m_{\rm hm}(y)|$ shown as a solid line. In the bulk $\mm\simeq 0$, and only the two nonzero boundary layers contribute to $\tilde{m}_1$ through:
\begin{equation}
\tilde{m}_1\simeq(w_1'(x)-w_2'(x))\int_0^\infty dy\, |m_{\rm hm}(y)|=-D_a\rho_b w'(x)/v\;. \label{eqn_int_m1}
\end{equation} 
Thus, eliminating $\rho_b$ in Eqs.~\eqref{eqn_boundary_m} and~\eqref{eqn_int_m1} gives 
\begin{align*}
    &m_x(x,w_{1}(x))=-\tilde{m}_1w_{1}'(x)/[w'(x)\ell_b]\;, \\
    &m_x(x,w_{2}(x))=\tilde{m}_1w_{2}'(x)/[w'(x)\ell_b]\;.
\end{align*}
The second line of Eq.~\eqref{eqn_tilde_m1} now reads 
\begin{equation*}
-D\partial_x\left[\tilde{m}_1\frac{w_1'^2(x)+w_2'^2(x)}{w'(x)\ell_b}\right]=\partial_x\left[\tilde{m}_1V_m'(x)\right]\;,
\end{equation*}
where
\begin{equation}
V_m(x)\equiv-\frac{D}{\ell_b}\int \frac{w_1'^2(x)+w_2'^2(x)}{w'(x)}\, dx\; \label{eqn_entropy_potential_m}
\end{equation}
is the effective potential that enters the dynamics of $\tilde
m_1$. We note that it differs from the
potential~\eqref{eqn_entropy_potential_inftyw} that enters the
effective dynamics of $\tilde\rho$.  We note that $V_m(x)$ depends
both on $w(x)$ and on the slopes of the two boundaries separately. If
the channel is symmetric such that $w'_2(x)=-w'_1(x)=w'(x)/2$, the
effective potential simplifies into
\begin{equation*}
V_m(x)=-\frac{Dw(x)}{2\ell_b}\;.
\end{equation*}

We note that the effective
potential~\eqref{eqn_entropy_potential_inftyw} entering the
dynamics of $\tilde{\rho}$ is unaltered in the presence of tilted flat
boundaries. Assuming that the wall is located at $y=0$, without loss of
generality, the density profile along $y$ in
Fig.~\ref{fig_corrections}a is given by the rescaling
\begin{equation}
  y\to y\cos\arctan w_1'(x)\sim y-y[w_1'(x)]^2/2
\end{equation}
in $\rho_{\rm hm}(y).$ The correction term scales as $[w_1'(x)]^2$,
which can be neglected at order $|w'(x)|$. Then, under the same
closure as in Sec.~\ref{sec_FJ_active_0th}, $\tilde{m}_2=0$, we obtain
an FJ-like set of equations for the density and orientation fields along the tube, valid up to order $|w'(x)|$,
\begin{align}
\partial_t \tilde{\rho}=&-v\partial_x\tilde{m}_1+D\partial_{xx}\tilde{\rho}+\partial_x\left[\tilde{\rho}V'(x)\right]\;. \label{eqn_tilde_rho_1st_trunc} \\
\partial_t \tilde{m}_1=&-(v/2)\partial_x\tilde{\rho}-\tilde{m}_1/\tau+D\partial_{xx}\tilde{m}_1+\partial_x\left[\tilde{m}_1V_m'(x)\right]\;.\label{eqn_tilde_m1_1st_trunc}
\end{align}
We note that $V_m(x)\neq V(x)$ so that a single conservative potential
cannot describe the 1d dynamics of 2d ABPs.

\subsection{Ratchet flows of ABPs along 2d corrugated channels}\label{sec_ratchet}
We now estimate the strength of the ratchet flow from
Eqs.~\eqref{eqn_tilde_rho_1st_trunc} and
\eqref{eqn_tilde_m1_1st_trunc}. To do so, we consider periodic tubes
with $w(0)=w(L)$, such that $\tilde{\rho}(0)=\tilde{\rho}(L)$ and
$\tilde{m}_1(0)=\tilde{m}_1(L)$. In the steady state,
$\partial_t\tilde{\rho}=\partial_t\tilde{m}_1=0$. Eq.~\eqref{eqn_tilde_rho_1st_trunc}
can be integrated once to obtain the flux of the particles $\tilde{J}$ as
\begin{equation}
\tilde{J}=v\tilde{m}_1(x)-D\tilde{\rho}'(x)-\tilde{\rho}(x)V'(x)\;. \label{eqn_tilde_J}
\end{equation}
We note that $\tilde{J}$ does not depend on $x$ in the steady
state. Integrating Eq.~\eqref{eqn_tilde_J} along the channel then leads to
\begin{equation}
\tilde{J}L=J_{\rm tot}[\tilde{\rho}]\equiv -\int_0^L \tilde{\rho}(x)V'(x)dx\;. \label{eqn_J_tot}
\end{equation}
The last equality comes from the periodic boundary conditions and the
fact that, in the steady-state, $\tilde{m}_1$ can be written as a
total derivative according to Eq.~\eqref{eqn_tilde_m1}. Note that
$J_{\rm tot}[\tilde{\rho}]=0$ whenever $\tilde{\rho}$ is a local
function of $V(x)$~\cite{OByrne_2022}. This applies, in particular, to
effective Boltzmann weights $\tilde{\rho}_B=\exp(-V(x)/D_{\rm eff})$.

We now compute the steady-state density $\tilde{\rho}_{\rm ss}$ and the
flux $\tilde{J}$ from Eqs.~\eqref{eqn_tilde_rho_1st_trunc} and
\eqref{eqn_tilde_m1_1st_trunc}. The steady state of
Eq.~\eqref{eqn_tilde_rho_1st_trunc} gives
\begin{equation*}
    \tilde{m}'_1(x)=\frac{D}{v}\tilde{\rho}''(x)+\frac{1}{v}\partial_x[\tilde{\rho}(x)V'(x)]\;.
\end{equation*}
Substituting it into Eq.~\eqref{eqn_tilde_m1_1st_trunc}, in the steady state we have
\begin{align}
    \tilde{m}_1=&\frac{\tau}{1-\tau V_m''}\left[D\partial_{xx}\tilde{m}_1+\tilde{m}_1'V_m'(x)-\frac{v}{2}\partial_x\tilde{\rho}\right] \nonumber \\
    =&\frac{\tau}{1-\tau V_m''}\left[(V_m'+D\partial_x)\left(\frac{D}{v}\tilde{\rho}''+\frac{1}{v}\partial_x(\tilde{\rho}V')\right)-\frac{v\tilde{\rho}'}{2}\right] \label{eqn_m1_ss}\\
    \simeq & \frac{\tau}{1-\tau V_m''}\left[\frac{V_m'\partial_x(\tilde{\rho}V')}{v}+\frac{D\partial_x(\tilde{\rho}V'')}{v}-\frac{v\tilde{\rho}'}{2}\right]\;. \label{eqn_m1_ss_approx}
\end{align}
From~\eqref{eqn_m1_ss} to~\eqref{eqn_m1_ss_approx}, we keep only terms up to the order $\tilde{\rho}'$. Substituting Eq.~\eqref{eqn_m1_ss_approx} into~\eqref{eqn_tilde_J} yields a first-order non-homogeneous linear ordinary differential equation of $\tilde{\rho}$. Its solution is given by the sum of the general solution $c \rho_h(x)$ of the complementary equation with $\tilde{J}=0$ and a particular solution with $\tilde{J}\neq 0$. $c$ is a constant real number that is set by the total number of active particles. We choose $c=\tilde{\rho}_{\rm ss}(0)$ for simplicity. First, $\rho_h(x)$ is given by:
\begin{align}
    \rho_h(x)= &\exp[-I(x)],\\
    I(x)\equiv &\int_0^x\frac{V'(l)-\tau [V_m'(l)V'(l)]'-D\tau V'''(l)}{\tilde{D}(l)}dl\;, \label{eqn_I_x}\\
        \tilde{D}(x)\equiv &D_{\rm eff}-\tau V_m'(x)V'(x)-D\tau[V_m''(x)+V''(x)]\;. \nonumber \\
\end{align}
We note that if $w(x)$ has parity symmetry, $V'$ and $V_m'$ are odd
functions, leading to $I(L)=0=I(0)$. The periodic boundary condition
holds, and $\tilde{J}=0$ is the proper solution of
Eq.~\eqref{eqn_tilde_J}. Thus no spontaneous currents appear in
channels with parity symmetry. We note that if $V_m=0$ and if
terms of order $w''(x)$ are neglected, $\tilde{D}\to D_{\rm eff}$ and
$I(x)\to V(x)/D_{\rm eff}$. $\rho_h(x)$ then reduces to the Boltzmann weight
Eq.~\eqref{eqn_general_boltzmann} with no spontaneous currents as
well.

If $w(x)$ does not have parity symmetry in $x$, $I(0)\neq I(L)$, and
$h(0)\neq h(L)$. Thus the periodic boundary condition requires
$\tilde{J}\neq 0$. The full solution of Eq.~\eqref{eqn_tilde_J} with
non-zero $\tilde{J}$ reads
\begin{equation*}
    \tilde{\rho}_{\rm ss}(x)=\rho_h(x)\left[\tilde{\rho}_{\rm ss}(0)-\tilde{J}\int_0^x \frac{[1-\tau V_m''(s)]\exp[I(s)]}{\tilde{D}(s)}ds\right]\;.
\end{equation*}
From the periodic boundary condition $\tilde{\rho}_{\rm ss}(0)=\tilde{\rho}_{\rm ss}(L)$, we can solve
\begin{equation}
    \tilde{J}=\frac{\tilde{\rho}_{\rm ss}(0)[\rho_h(L)-1]}{\rho_h(L)}\left[\int_0^L \frac{[1-\tau V_m''(s)]\exp[I(s)]}{\tilde{D}(s)}ds\right]^{-1}\;. \label{eqn_tilde_J_sol}
\end{equation}
We note again that a non-local functional dependence of $w(x)$ in Eq.~\eqref{eqn_I_x} is necessary to give a non-zero $\tilde{J}$.

Finally, we check the theoretic prediction of $\tilde{J}$ in
agent-based simulations (Fig.~\ref{fig_ratchet_flow}). In the
simulations, we use asymmetric tubes with $w(x)=A(1-\cos(2\pi
(x/L)^N))+2C_0$. With increasing $N$, the maximum of $w'(x)$
increases, and the strength of the ratchet current $J_{\rm tot}$
increases, which we measure from simulations as $J_{\rm
  tot}=\langle\dot{x}(t)\rangle_t$. We find the analytical prediction
$\tilde{J}L$ with $\tilde{J}$ given by Eq.~\eqref{eqn_tilde_J_sol}
gives predictions of order-of-magnitude agreements with measurements
from simulations. We note that neglecting $V_m$ in
Eq.~\eqref{eqn_tilde_J_sol} leads to no ratchet flow. We also note
that, given $\tilde{\rho}_{\rm ss}(x)$ measured from simulations, the
value of $J_{\rm tot}[\tilde{\rho}_{\rm ss}]$ using the generalized
entropy potential given in Eq.~\eqref{eqn_entropy_potential_inftyw}
also agrees with the measurements in simulations.

\section{Discussion}\label{sec_discussion}

In this article, we extended the Fick-Jacobs equation from passive
particles to active particles in the limit of a wide tube
$w(x)\gg\ell_b$ with slow varying width $|w'(x)|\ll 1$. Under such
assumptions, the local translational symmetry along the boundary can
be recovered and the boundary layers decouple. Contributions to the
integration of the probability density of particles from the bulk and
boundaries can be separated. We considered different levels of
approximations. With the assumption of locally homogeneous boundaries,
we constructed a generalized entropy potential $V(x)$
Eq.~\eqref{eqn_entropy_potential_inftyw} in which the 1d density of
active particles $\tilde{\rho}$ evolves. The approximation of locally
flat and horizontal boundaries captures well the dynamics in large
spatial-temporal scales in a long tube. The steady state predicts the
density profiles better than that from the passive entropy potential
using an effective temperature given by the large-scale diffusivity of
the active particles. Using the Kramers' law, we can predict the
mean-first-passage time with the generalized effective potential
Eq.~\eqref{eqn_entropy_potential_inftyw}. The analysis for ABPs
automatically holds for RTPs, because ABPs and RTPs share the same
dynamics of the density and polarization fields~\cite{Cates_2013,
  Solon_2015}. The analysis can be extended to higher dimensions as
well. Note that the calculations require a finite $D$, which ensures a
continuous density profile near the boundary and a nonzero $\ell_b$.

We then worked beyond the assumption of vanishing $w'(x)$ and allowed
the boundary to be tilted. This led to a correction term in the
dynamics of the orientation field $\tilde{m}_1$, which amounts to a
drift term stemming from a different entropy potential $V_m(x)$. While
the improvement in the prediction of the density profiles is not
significant at large scales, this term enables us to predict the
spontaneous ratchet flow observed in simulations of active particles
in asymmetric channels. We note that $V(x)\neq V_m(x)$ so that the
problem does not reduce to 1d dynamics of ABPs in a single
conservative potential.

Note that further analytical progress could be made by truncating at
higher order moments and by considering the curvature of the
boundaries. This would require calculating the probability density
near a curved boundary~\cite{Wagner_2022}. How to account for the deviation from the
local steady state in the $y$-direction at finite $w'(x)$ remains at
this stage a challenging question.

Then, in this work, dilute spherical particles are
considered. Elongated active particles can align with rigid boundaries
and accumulate as well~\cite{Elgeti_2009}. The probability of
particles leaving the boundaries then decreases with a larger aspect
ratio, and the number of accumulated particles increases with the
length of the particle in the dilute limit~\cite{Elgeti_2009}. A
larger boundary accumulation is expected to further flatten the
density profile $\tilde{\rho}$ of elongated particles. For dense
systems, the interactions between particles will also impact the
boundary layer close to the wall. 

Finally, we note that the experimental setup considered here could be
easily engineered for many type of synthetic active particles ranging
from Quincke rollers~\cite{bricard2013emergence} to phoretic Janus
colloids~\cite{theurkauff2012dynamic}.

\if{
Finally, possible relevant experimental systems can be colloid
particles driven by an external magnetic field moving in PDMS
microfluid channels if the contribution from hydrodynamics of the
water can be neglected and the boundaries can be considered rigid
walls. Robots moving on a track with varying widths could be a clean
experimental system as well.}\fi

\begin{acknowledgments}
The author thanks Xiaqing Shi, Hepeng Zhang, Hugues Chat\'{e}, and
Masaki Sano for helpful discussions and critical comments. Special
thanks to Julien Tailleur for helping with the writing of this
manuscript. The author acknowledges support from the start-up grant
from Soochow University.
\end{acknowledgments}

\bibliography{biblio}

\begin{thebibliography}{45}
\bibitem [{not()}]{note_oneforth}%
  \BibitemOpen
  \href@noop {} {}\bibinfo {note} {A factor of $1/4$ appears because the
  particles can escape from two sides, and we only calculate the time ending at
  the peak of the entropy potential when particles have probability $1/2$ to
  escape.}\BibitemShut {Stop}%
\end{thebibliography}

\end{document}